**TOURISTS' DIGITAL FOOTPRINT IN CITIES: COMPARING BIG DATA SOURCES**

MARÍA HENAR SALAS-OLMEDO *
Departamento de Geografía Humana.
Universidad Complutense de Madrid
C/Profesor Aranguren, s/n. 28040 Madrid, Spain
msalas01@ucm.es

JUAN CARLOS GARCÍA-PALOMARES
Departamento de Geografía Humana.
Universidad Complutense de Madrid
C/Profesor Aranguren, s/n. 28040 Madrid, Spain
jcgarcia@ucm.es

JAVIER GUTIÉRREZ
Departamento de Geografía Humana.
Universidad Complutense de Madrid
C/Profesor Aranguren, s/n. 28040 Madrid, Spain
javiergutierrez@ghis.ucm.es

*Corresponding author.

# TOURISTS' DIGITAL FOOTPRINT IN CITIES: COMPARING BIG DATA SOURCES

**Abstract**.- There is little knowledge available on the spatial behaviour of urban tourists, and yet tourists generate an enormous quantity of data (Big Data) when they visit cities. These data sources can be used to track their presence through their activities. The aim of this paper is to analyse the digital footprint of urban tourists through Big Data. Unlike other papers that use a single data source, this article examines three sources of data to reflect different tourism activities in cities: Panoramio (sightseeing), Foursquare (consumption), and Twitter (being connected). Tourist density in the three data sources is compared via maps, correlation analysis (OLS) and spatial self-correlation analysis (Global Moran's I statistic and LISA). Finally the data are integrated using cluster analysis and combining the spatial clusters identified in the LISA analysis in the different data sources. The results show that the data from the three activities are partly spatially redundant and partly complementary, and allow the characterisation of multifunction tourist spaces (with several activities) and spaces specialising in one or various activities (for example, sightseeing and consumption). The case study analysed (Madrid) reveals a significant presence of tourists in the city centre, and increasing specialisation from the centre outwards towards the periphery. The main conclusion of the paper is that it is not sufficient to use one data source to analyse the presence of tourists in cities; several must be used in a complementary manner.

**Key words**.- Urban tourism; Big data; Photo-sharing services; Social networks; Spatial analysis; GIS

## 1. INTRODUCTION

Big Data is a new concept that has become widely popularised in recent years to describe the production of massive quantities of data. Big Data covers a range of very different areas: Internet searches, GPS logs, bank card transactions, records of mobile phone activity, social networks, data on water and electricity consumption, meteorological data, images recorded with video cameras and many more. The main characteristics of these new data sources include particularly the following three Vs: *volume*, in terabytes or petabytes of data; *velocity*, created in or at near real time; and *variety*, taken from a wide variety of sources, either structured (data that can be stored in the form of tables), semistructured (htlm files) or unstructured (texts, photographs, videos) (McAfee et al., 2012; Kitchin, 2013; Sagiroglu and Simanc, 2014).

One of the fields in which Big Data offers the greatest opportunities is tourism. Official data sources do not provide detailed information on the places tourists visit in cities. At best, they offer figures derived from surveys, hotel records or museum admissions. Big Data supplies a large quantity of information to complement the traditional sources. Tourists leave a digital "footprint" in most of their activities, and these new data sources now make it possible to analyse tourists' behaviour in the cities they visit. They take vast numbers of photographs and upload them to photo-sharing services, they make payments with bank cards, they talk and send messages via their mobile phones, they are active on social networks, and so on. All this activity produces an enormous quantity of digital data (Big Data) which can be analysed to study behaviour patterns (Shoval and Isaacson, 2007; Asakura and Iryo, 2007; Girardin et al., 2008a and b). Much of these data is geolocated, so tourists' activity can be analysed spatially. However, there are very few papers that apply Big Data to examine the spatial distribution of tourists in cities, probably due to the novelty of these information sources and the fact that some are difficult to access.

Photo-sharing services provide very useful information for identifying the presence of tourists when they go sightseeing in cities. Although there are several photo-sharing communities that allow the geolocation of photos (such as Flickr or Instagram), Panoramio is probably the most useful service for measuring tourist hotspots, as this website shows photographs taken of places or landscapes when sightseeing, which are then posted online once they have been georeferenced. The records of geolocated photographs can be used not only to identify sightseeing spots (García-Palomares et al., 2015), but also to analyse the spatial and temporal patterns of tourist flows in cities Girardin et al., 2008b).

However, tourists not only visit the most photographed spaces. They also go shopping, go to restaurants and stay in hotels, and they leave their digital footprint in all these establishments when they pay with a bank card or check-in their location on social networks. These digital footprints of tourists offer information which is largely complementary to the data from photo-sharing social networks. The most photographed areas often have very little offer of accommodation and shopping. In the case of business tourists, their hotel and the spaces they frequent tend to be close to business sectors, and not necessarily in the most photographed areas.

During their stay in the city, tourists also log onto the Internet to confirm details of their visit, check their e-mail, engage in the social networks, and so on. This activity also leaves a digital footprint in many of the places they visit. Tourists often use the facilities in hotels, hostels, restaurants and certain open spaces to connect to Internet through free WiFi networks, so their activity on social networks may particularly reflect this type of spaces.

The main aim of this paper is to compare three geolocated data sources to identify the presence of tourists in cities in terms of their different activities: sightseeing, consumption and being connected. The sightseeing activity is analysed though data from photo-sharing services, specifically from the platform Panoramio. Consumption is tracked by Foursquare check-ins. Finally, Internet activity is measured based on data on the social network Twitter. The study area is the city of Madrid, one of the European cities with the highest volume of tourists.

This paper contributes to the literature on Big Data and tourism activity from a threefold perspective: 1) Three different data sources are compared to obtain the most comprehensive analysis of tourists' location. 2) The data for tourist activity (photos, check-ins, tweets,) are not analysed directly as in previous works, but the tourists themselves are the unit of analysis. The data are processed to allow the number of tourists to be counted in each place in the city according to each data source, thus avoiding problems of multiple counting, and making the results comparable. 3) The information from the different data sources is integrated through cluster analysis and spatial autocorrelation analysis to characterise the areas of tourist concentration according to the type of activity.

The remainder of this paper is structured as follows. Section 2 summarises the existing literature on the use of photo-sharing services, Foursquare check-ins and Twitter in urban studies, with a particular focus on tourism. Sections 3 and 4 describe the data and the methodology respectively. Section 5 describes and discusses the results, while Section 6 presents the main conclusions and suggests further directions for research.

## 2. RELATED LITERATURE

### Sightseeing: photo-sharing services

Sightseeing is one of the main tourist activities in cities, and leaves its digital footprint on social networks for photo-sharing such as Instagram, Flickr and Panoramio. All three offer the possibility of geolocating photographs, but Panoramio (http://www.panoramio.com) particularly allowed the georeferencing of the photos, as it focused on images of places or landscapes shared by its users, which can be seen on the Panoramio website (until November 4, 2016) or through Google Earth and Google Maps. In fact, Panoramio was a Google service, and had over 120 million geolocated photographs (2015 data according to Panoramio data API).

Photo-sharing services have been used for several purposes in the field of tourism, including identifying social events such as festivals, demonstrations, sporting events and so on (Sun and Fan, 2014), estimating tourist numbers (Koerbitz et al., 2013), identifying the presence of tourists (Girardin et al., 2008a, Kisilevich et al., 2013; Straumann et al., 2014) and the most common trajectories followed by tourists (Girardin et al., 2008b), proposing or assessing tourist routes (Kurashima et al., 2013), suggesting tourist trips (Lu et al., 2010) and planning trips lasting several days and places to visit (Li, 2013). These data sources make it possible to identify areas with a concentration of tourists in cities (sightseeing spots) through spatial statistical techniques (García-Palomares et al., 2015). Photographs taken by tourists can be differentiated from those taken by residents based on the time period in which the same user takes the photographs. The results obtained in the work of García-Palomares et al. (2015) show clearly differentiated spatial distributions for tourists (more concentrated in sightseeing spots) and local residents (more widely dispersed throughout the city).

**Consumption: Foursquare check-ins**

Tourists not only leave their digital footprint when they go sightseeing in the city. They also leave a digital trail when they engage in consumption-related activities (shopping, restaurants and so on) and pay with bank cards (Sobolevsky et al., 2014a, 2014b and 2015). However, bank card transactions are a data source that has been very little used in tourist studies, due to the difficulty of accessing these databases. An alternative data source for analysing consumption activities is Foursquare social network. This service enables a user to inform their friends his/her location by generating check-ins, rate and review venues they visit, and read other users' reviews. Each check-in contains information that reveals who (which user) spends time where (at what location), when (what time of day, what day of week), and doing what (according to the kind of venue) (Çelikten et al., 2016).

Foursquare data have been used for analysing geographic distribution of venues across the cities (Çelikten et al., 2016), discovering functional urban areas (Vaca et al., 2015), analysing movement pattern and the popularity of areas in cities (Silva et al., 2013), studying traffic conditions (Ribeiro et al, 2014), and performing trade area analysis (Qu et al., 2013). However, there are few studies using Foursquare data in the field of tourism. One exception is Ferreira et al. (2014), who analyze the spatio-temporal characteristics of the behavior of tourists and residents in a set of cities, identifying the most visited places and the temporal distributions of both groups of visitors. Other exception is the paper by Serrano-Estrada et al. (2016), who map the most relevant venues in terms of number of visitors within a tourist city.

**Being connected: Twitter**

Most of the studies done with mass data from social networks have used Twitter (Murthy, 2013), not only because this platform has global coverage, but also because its data (the tweets) are available free on the Internet the instant they are produced –that is, in real-time. Each geolocated tweet leaves the digital footprint of the place and the time it was sent. If the data are processed according to their user identifier, they provide an approximation of the places visited by each user at different times of the day and days of the week –that is, their spatial and temporal profile.

Activity on social networks can be used as a proxy to analyse the changing population densities in the city throughout the day (Ciuccarelli et al., 2014) and the population's mobility patterns (Wu et al., 2014, Salas-Olmedo and Rojas-Quezada, 2016). Twitter's daily use profiles serve to classify the space according to the type of dominant activity, whether business, leisure/weekend, nightlife and residential (Frias-Martinez et al, 2012). Geolocated tweets have also been used to analyse the degree of social mixing in the use of space by tracking the movement of social groups in highly segregated cities such as Río de Janeiro (Netto et al., 2005) and Louisville (Shelton et al., 2015). Unlike the information supplied by official sources, which offer data relating to place of residence, these studies apply indicators of multiculturalness and mixing to examine the use of space throughout the day. For example, there are studies on linguistic diversity in cities and regions based on the language used in tweets as an indicator of cultural diversity (Mocanu et al., 2013). Elsewhere, Takhteyev et al. (2012) studied the role of distance, languages and borders in configuring the networks of tweets, and concluded that the relations established over long distances are similar to those observed in air transport flows.

Papers on tourism using Twitter data are particularly scant. The very few works that use geolocated tweets in the field of tourism tend to focus on comparing visitor's spatial behaviour between cities on the national or global scale (Bassolas et al., 2016; Hawelka et al., 2014; Sobolevsky et al., 2015), but do not address the spatial patterns within the city.

## 3. DESCRIBING AND PRE-PROCESSING THE DATA

### Panoramio

The records of 307,062 geolocated photographs uploaded to Panoramio between 2006 and 2014 covering the municipality of Madrid were downloaded from the Panoramio API. This dataset contains information about the geographic coordinates, the ID of the owner of the photograph, a URL link to the photograph, and the date on which it was uploaded (day, month and year).

This downloading generated ".csv" files. The geographical coordinates of each photograph were used to create a layer for each location using ArcGIS 10.4. This data source does not include a field to record the user's nationality. However, it was possible to "identify" each user as a tourist or a resident using the user's ID and the date of his/her photograph. Using MongoDB, if a user had taken pictures in Madrid over a period exceeding one week per year, the photographs were attributed to residents; if the period was less than one week per year, then they were attributed to tourists. This methodology is similar to that used by Fischer for his Geotaggers' World Atlas, and García-Palomares et al. (2015). A total of 27,573 photographs were assigned to tourists as a result of this process (Figure 1a).

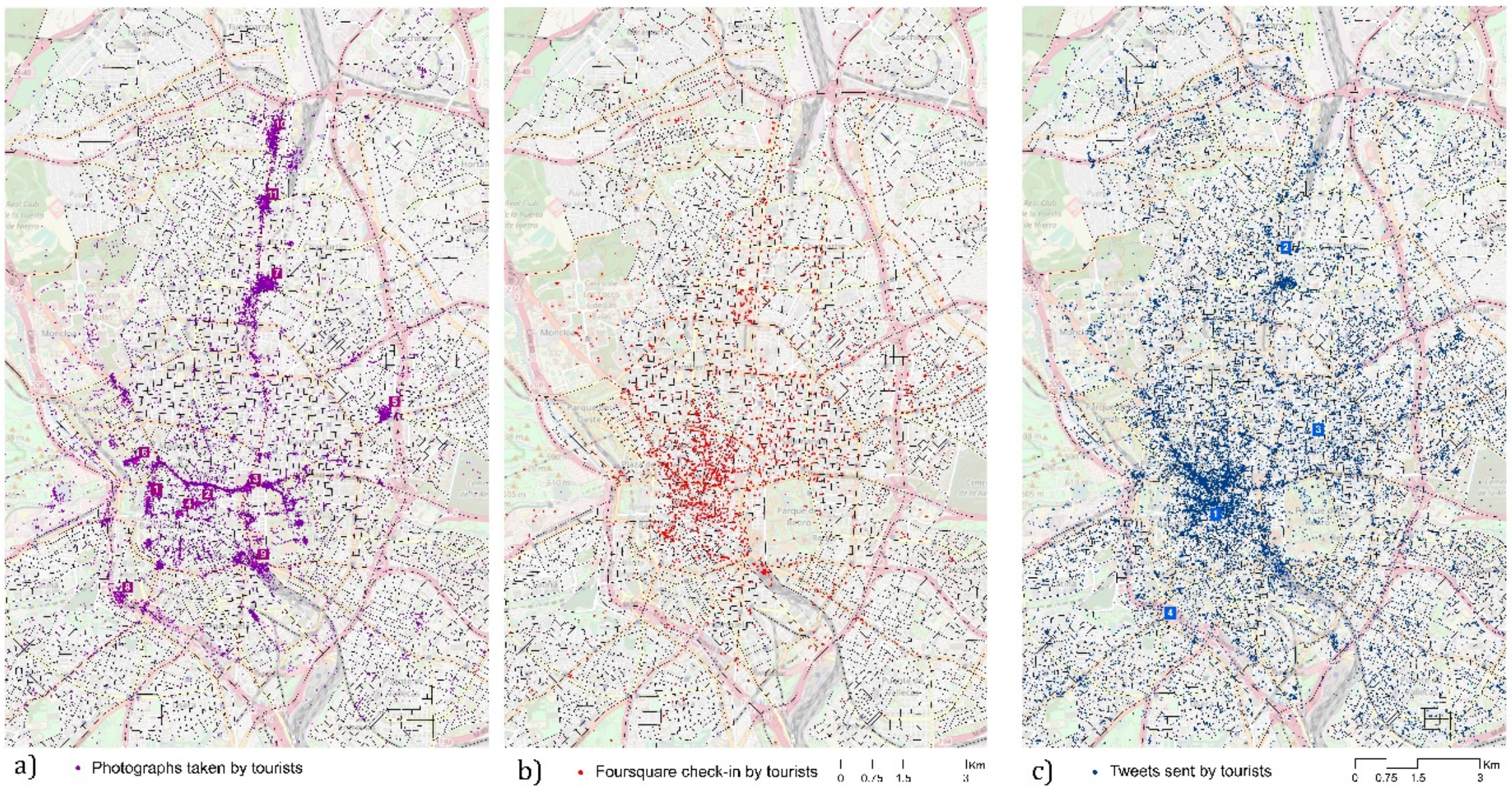

Figure 1: a) Photographs taken by tourists. b) Foursquare check-in by tourists. c) Tweets sent by tourists.

*Figure 1a references*: 1) Royal Palace; 2) Puerta del Sol square; 3) Plaza de Cibeles square; 4) Plaza Mayor square; 5) Las Ventas bullring; 6) Temple of Debod; 7) Real Madrid football stadium; 8) Atlético de Madrid football stadium; 9) Atocha – Reina Sofía Museum; 10) Cuatro Torres; 11) Torres Kio
*Figure 1c references*: 1) Historic centre; 2) Salamanca district; 3) Paseo de la Castellana axis; 4) Madrid Río axis

*Background map:* OpenStreetMap.

**Twitter and Foursquare**

One of the most widely used social networks is Twitter, a platform that allows users to send messages with a maximum of 140 characters –known as tweets– which by default are public. This service has 500 million users all over the world and generates around 65 million tweets a day. Some of these (around 3% until April 2015 and currently around 1%) are geolocated; that is, messages in which the sender's location is known from their geographic coordinates. Given the vast amount of messages sent on the social networks, this 3% percentage represents an enormous quantity of tweets.

Our study was conducted using a dataset of geolocated tweets sent from Madrid between 2012 and 2014. The dataset contains information on the geographic coordinates, the owner's ID, the language, the date on which it was uploaded (minute, hour, day, month and year), any messages included, etc. This file underwent a similar treatment to the Panoramio file. First, tourists were identified using MongoDB to extract users that had tweeted for a week or less per year. Then, all twees from those users within the municipality of Madrid were used to generate a layer of tweets with ArcGIS using the coordinates of the logs. A total of 234,159 tweets were identified as sent by tourists visiting Madrid.

Foursquare is a community of 50 million users allowing tourists to check in to the establishments they visit and providing personalized recommendations of the places to go (restaurants, nightlife spots, shops and other places of interest) in the surrounding area. These check-ins are private by default, but they become publicly accessible for example, when users opt to share their check-ins publicly via Twitter (Çelikten et al., 2016). Thus, it was possible to extract Foursquare data by selecting Foursquare check-ins in our Twitter dataset. From a total of 234,159 tweets assigned to tourists, 20,076 were Foursquare check-ins (Figure 1b) and the rest (214,083) were considered ordinary tweets (Figure 1c).

Foursquare data and tweets exhibit very different temporal distribution patterns (Figure 2). Foursquare check-ins are more equally distributed along the day (check-ins in restaurants, shops and other places), while tweets tend to be more concentrated between 18 and 21 hours (what suggests that the use of Twitter is predominantly related to the location of accommodations).

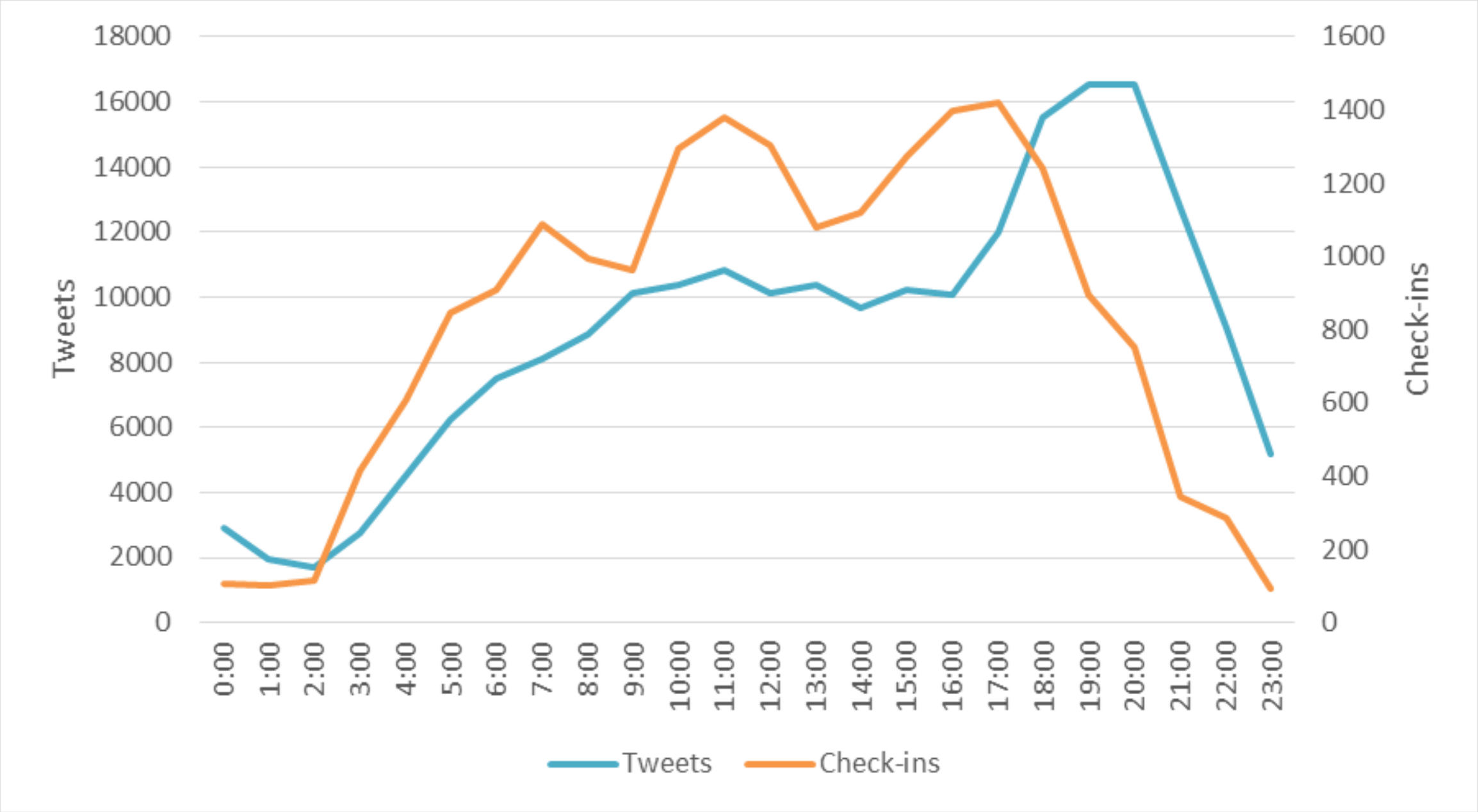

Figure 2: Temporal distribution of Foursquare and Twitter data of tourists in Madrid

## 4. METHODOLOGY

The following methodology was used to analyse the spatial distribution of tourists in Madrid:

a) Number of tourists per census tract.- The tourists' photos, Foursquare check-ins and tweets were located according to census tracts, counting the number of single tourists in each census tract and for each data source from the user ID. The row data was converted to single tourists using joint spatial (census tracts) aggregation to obtain the number of single users in each census tract.

b) Tourist density by census tract.- The number of tourists in each census tract depends on the actual concentration of tourists in the census tract and its size. Larger census tracts tend to register a greater number of tourists. To mitigate this problem (Modifiable Areal Unit Problem -MAUP), the tourist density per census tract was obtained for each data source.

c) Rescaling of the data.- A very different number of tourists was detected through the three data sources. The density data of visitors by census tract were rescaled to a scale of 0 to 1,000 (by means of a linear transaction) in order to eliminate the effect of different ranges in the data sources and to make density data comparable.

d) Density maps and descriptive statistics.- Rescaled data aggregated by census tract were used to obtain density maps and descriptive statistics. Density maps provided an initial visual overview of the density distribution of tourists in Madrid according to the three datasets. Descriptive statistics were useful to compare the degree of concentration of visitors according to the data sources used.

e) OLS analysis.- Results obtained for the three data sources were compared using bivariant Ordinary Least Squares (OLS). The coefficient of determination reveals the common part of variation between each pair of data sources, and the differences between each pair of sources can be analysed using the maps of standardised residuals.

f) Cluster analysis.- Cluster analysis was used to integrate the information from the three data sources and characterise the census tracts according to the tourist activities performed in them. The K-means clustering algorithm was used. This method looks for a solution where all the features within each group are as similar as possible, and all the groups themselves are as different as possible.

g) Spatial autocorrelation analysis.- Unlike previous analyses, spatial autocorrelation techniques do not consider each location in an isolated way, but in relation to the locations in its environment (Anselin, 1995). Global Moran's I statistic and Anselin Local Moran's I (LISA statistic) were calculated separately for the three data sources, using the IDW (inverse distance weighted) method with a 500 m radius. The LISA analysis identified High-High clusters of tourists (a high value surrounded primarily by high values). The results were combined in order to determine the specialisation of each census tract, which can be classified as areas of sightseeing (Panoramio High-High clusters), tourist consumption (Foursquare High-High clusters), Internet consumption (Twitter High-High clusters) or a combination of two or three types (for example, consumption and sightseeing ). A census tract classified as HH cluster according to the three data sources means that in a radius of 500 m there is high potential for the three types of activities.

All these calculations and maps were made using ArcGIS 10.4 software.

## 5. RESULTS

### Tourist density maps and descriptive statistics

Figure 3 shows tourist density maps according to census tracts, with rescaled data and at the same intervals in the three maps. Panoramio (Figure 3a) clearly shows a high spatial concentration of visitors in the historic centre and along the north-south axis of the city (Paseo de La Castellana). The areas of greatest density denote the sightseeing spots in the city, for example Plaza de Cibeles, Puerta de Alcalá, Puerta del Sol, Plaza Mayor, the Royal Palace, the Temple of Debod, Plaza de España, Reina Sofía Museum, Atocha Station, Gran Vía street, the Real Madrid and Atletico de Madrid stadiums, the Las Ventas bullring, Torres Kio, Cuatro Torres, among others. Some census tracts have a high density of photos, not because they themselves contain a sightseeing spot, but because they offer a good vantage point for taking photographs of a sightseeing spot located in an adjacent census tract. The prohibition against taking photographs inside some monuments explains the relatively low density of photos in the census tracts containing them (for example, the Prado Museum and the Royal Palace).

The digital footprint of tourists in consumption activities (Figure 3b) is particularly dense in the historic centre and, to a lesser extent, in surrounding areas (for example, the area of luxury shops known as the Golden Mile in Barrio de Salamanca). It is also sporadically explained by the presence of some shopping centres (such as the AZCA centre) or singular places (Real Madrid Stadium).

Finally, tourist density identified according to Twitter (Figure 3c) is more disperse. It is particularly high in the historic centre and along the Paseo de la Castellana axis, and it tends to spread throughout a great number of census tracts.

The descriptive statistics of tourist density according to the data sources (Table 1) show that a much higher number of tourists have been detected from Twitter than from any of the other two sources, thus justifying the rescaling of the three variables. The coefficient of variation of Foursquare data reveals that, as suggested by the maps, consumption activities exhibits the highest spatial concentration.

Table 1: Descriptive statistics

| Tourist density/Ha according to census tracts | | | |
|---|---|---|---|
| | Panoramio | Foursquare | Twitter |
| Count: | 2415 | 2415 | 2415 |
| Minimum: | 0 | 0 | 0 |
| Maximum: | 3.83 | 8.25 | 13.86 |
| Sum: | 312.95 | 424.53 | 2491.02 |
| Mean: | 0.13 | 0.18 | 1.03 |
| Standard Deviation: | 0.33 | 0.55 | 1.31 |
| Variation coefficient | 250.83 | 305.56 | 127.02 |
| Tourist density/Ha according to census tracts: rescaled data | | | |
| | Panoramio | Foursquare | Twitter |
| Count: | 2415 | 2415 | 2415 |
| Minimum: | 0 | 0 | 0 |
| Maximum: | 1000 | 1000 | 1000 |
| Sum: | 81625.85 | 51420.18 | 179479.38 |
| Mean: | 33.80 | 21.29 | 74.32 |
| Standard Deviation: | 84.86 | 66.60 | 94.51 |
| Variation coefficient | 251.07 | 312.80 | 127.17 |

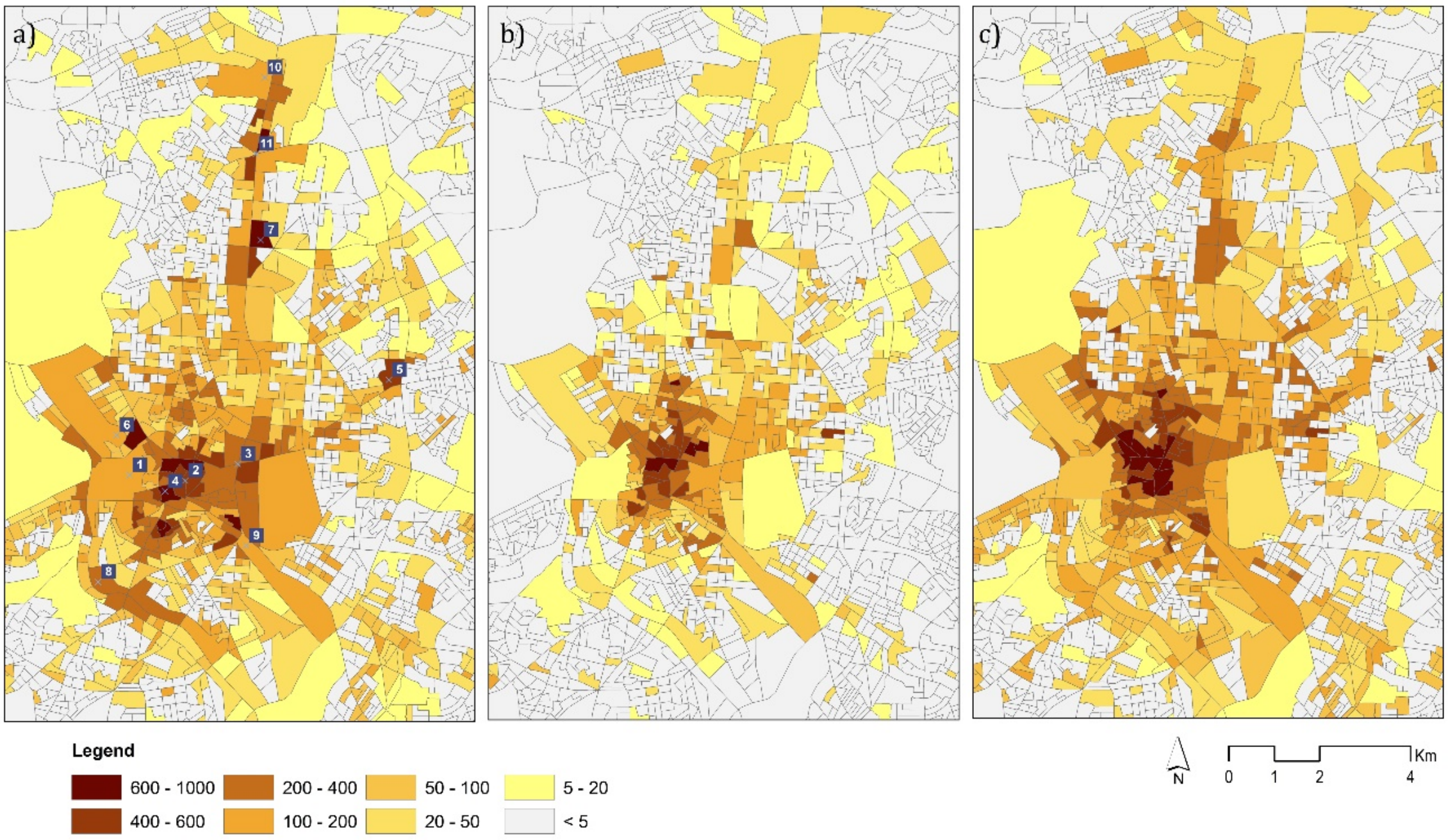

Figure 3: Tourist density according to their activities: a) sightseeing (Panoramio), b) consumption (Foursquare), c) being connected (Twitter). The references to the numbers in Figure 3a can be seen in Figure 1.

## Comparison between data sources: OLS analysis

To determine the degree of association between the three distributions, the coefficient of determination was calculated between each pair of data sources (Table 2). The standardised residuals of the bivariate regressions reveal where the data sources present the greatest differences (Figure 3).

The correlation analysis points to a medium positive correlation between the tourist densities provided by the Twitter-Foursquare data sources. The correlation between the number of tourists provided by the Panoramio-Twitter and Panoramio-Foursquare data sources is low, indicating a greater complementarity. The main discrepancies between these three data sources can be analysed through the residuals between the variables in the regression analysis:

- Panoramio has more tourists than expected according to Foursquare and Twitter (positive residuals in Figures 4a and 4b) in the historic centre and in the main sightseeing spots in the city (football stadiums, the bullring, Retiro Park, Torres Kio, Cuatro Torres, Temple of Debod), but less than expected (negative residuals) around the city centre.

- Foursquare shows more tourists than expected according to Twitter (Figure 4c) in the historic centre, the Salamanca district, in the AZCA shopping and business centre, the large shopping centre of La Vaguada, Real Madrid Stadium, and Atocha station, but less tourists than expected in less central areas.

Table 2: Coefficients of determination (adjusted $r^2$) of tourist distribution according to data sources (OLS)

| | Panoramio | Foursquare | Twitter |
|---|---|---|---|
| Panoramio | 1 | | |
| Foursquare | 0.27** | 1 | |
| Twitter | 0.31** | 0.52** | 1 |

** Significant to 0.01

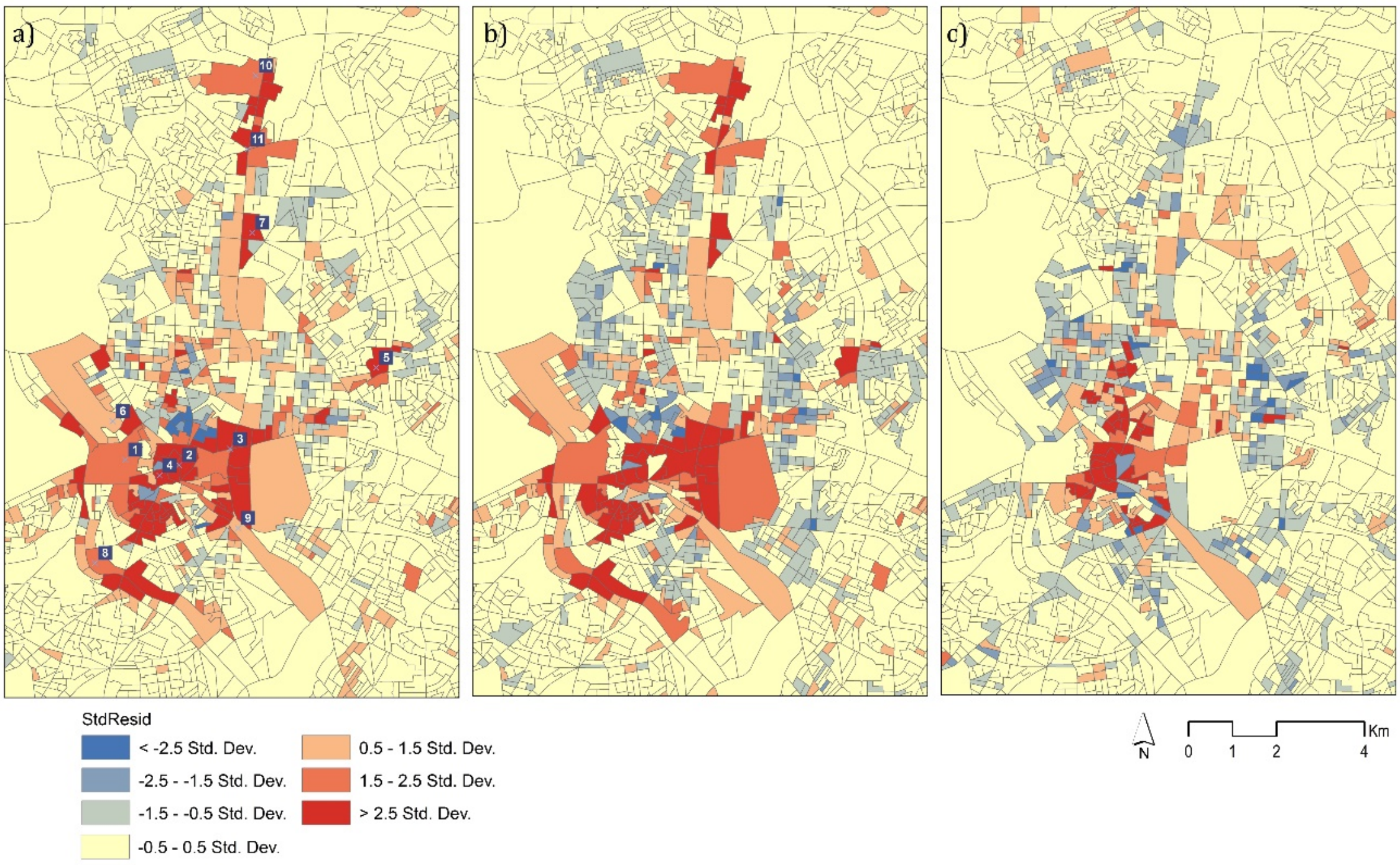

Figure 4: Standardised residuals of bivariate regressions: a) Foursquare-Panoramio; b) Twitter-Panoramio; c) Twitter- Foursquare. The references to the numbers in Figure 4a can be seen in Figure 1.

**Types of spaces according to tourist activities: cluster analysis**

Cluster analysis enables the census tracts to be typified according to tourist density per activity (sightseeing, consumption or Internet activity). The cluster was calculated using the K-means algorithm and using rescaled density values. Table 3 shows the means and standard deviations that characterise each group and the parallel box plot graph summarises both the groups and the variables within them (Figure 5). Figure 6 shows the spatial distribution of the 6 groups established:

- *Groups of census tracts with a predominance of tourists related to sightseeing*: groups 6 (coffee) and 2 (red). The differences are based on the intensity of the number of tourists. Group 6 contains tracts with a very high number of tourists identified in Panoramio, and high in Twitter and Foursquare users, and corresponds to spaces such as the bullring, Puerta de Alcalá arch, Glorieta de Atocha and Plaza de España square. Group 2 also contains tracts linked to sightseeing, but with a lower number of tourists: the parks of the Temple of Debod and Retiro, the Royal Palace and the surrounding area, the Atlético de Madrid stadium, Atocha station and singular buildings such as Torres Kio and Cuatro Torres.
- *Groups of census tracts with a predominance of tourists linked to consumption*: groups 5 (violet) and 3 (green). Group 5 contains the most commercial tracts in the historic centre (Gran Vía, Puerta del Sol square), which also have a very high number of tourists according to Twitter and Panoramio. Group 3 comprises tracts with a high tourist density related to consumer activities and Twitter, but much less in relation to Panoramio. These correspond to the historic centre (except the most commercial area), with a decrease in specialisation in Twitter towards the periphery, but also commercial areas like Golden Mile-Salamanca district, AZCA and retail spaces in the centre.

- Groups of *census tracts with a predominance of tourists related to Internet activities*: Group 1 (blue), with a lower number of tourists, occupies the limits of the historic centre.
- *Census tracts with a low presence of tourists*: group 4 (yellow). These are the non-tourist spaces in the city and generally correspond to peripheral census tracts.

Table 3: Means and standard deviations of the groups of tracts.

| | | Tourists according to Panoramio | | Tourists according to Foursquare | | Tourists according to Twitter | |
|---|---|---|---|---|---|---|---|
| Groups | Number of tracts | Mean | Sd | Mean | Sd | Mean | Sd |
| 1 | 331 | 23.9 | 30.0 | 33.2 | 35.1 | 161.3 | 60.1 |
| 2 | 151 | 148.47 | 56.0 | 21.4 | 27.7 | 110.5 | 66.1 |
| 3 | 76 | 141.5 | 82.9 | 182.6 | 72.5 | 282.6 | 102.1 |
| 4 | 1795 | 8.2 | 17.5 | 3.82 | 9.2 | 36.1 | 25.2 |
| 5 | 26 | 343.0 | 217.3 | 498.7 | 165.1 | 599.9 | 178.1 |
| 6 | 36 | 466.6 | 168.2 | 96.9 | 78.8 | 208.9 | 119.6 |
| Total | 2415 | 33.8 | 84.8 | 21.3 | 66.6 | 74.3 | 94.5 |

-

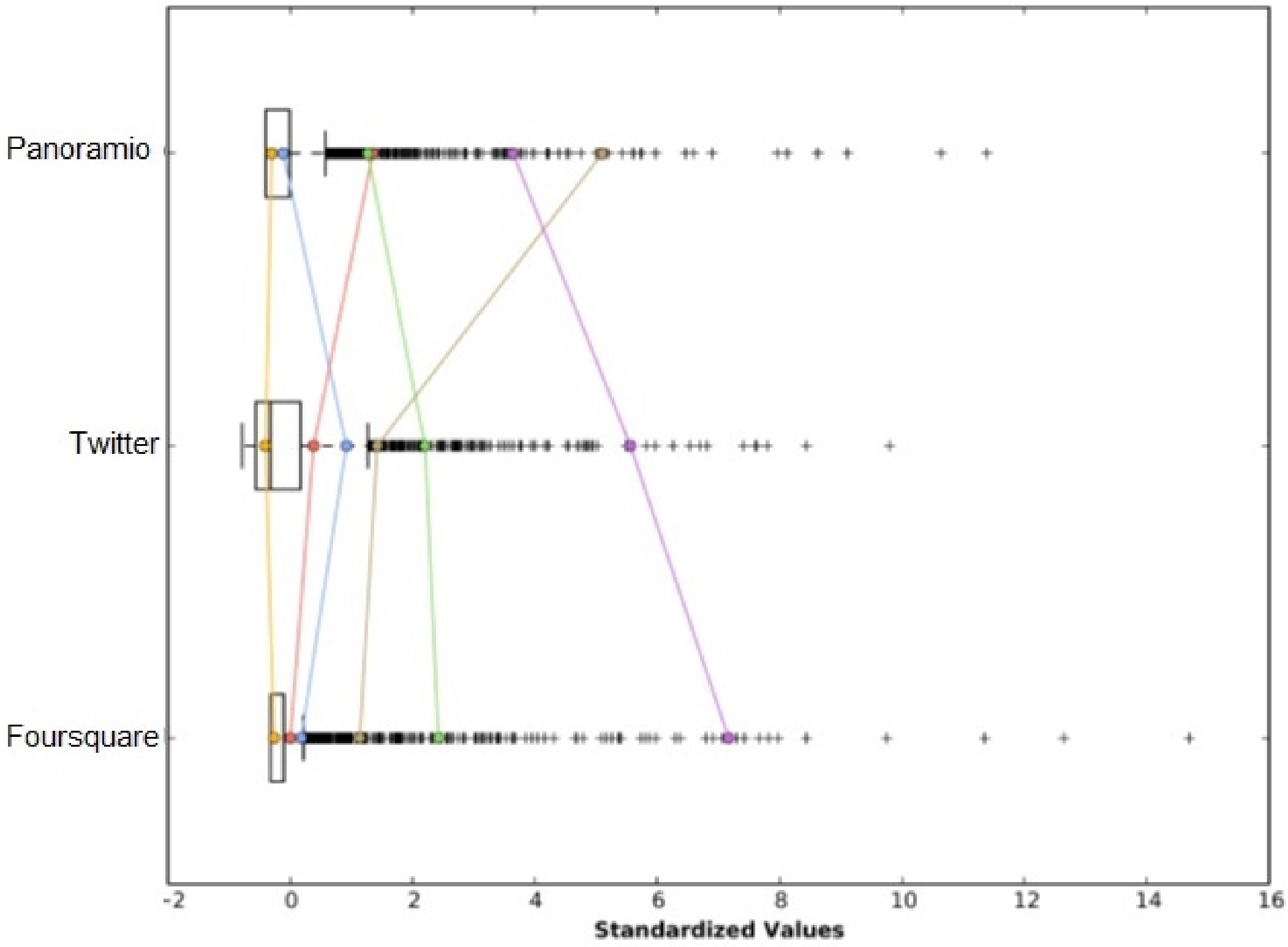

Figure 5: Standardised means for the groups of tracts (parallel box plot graph).

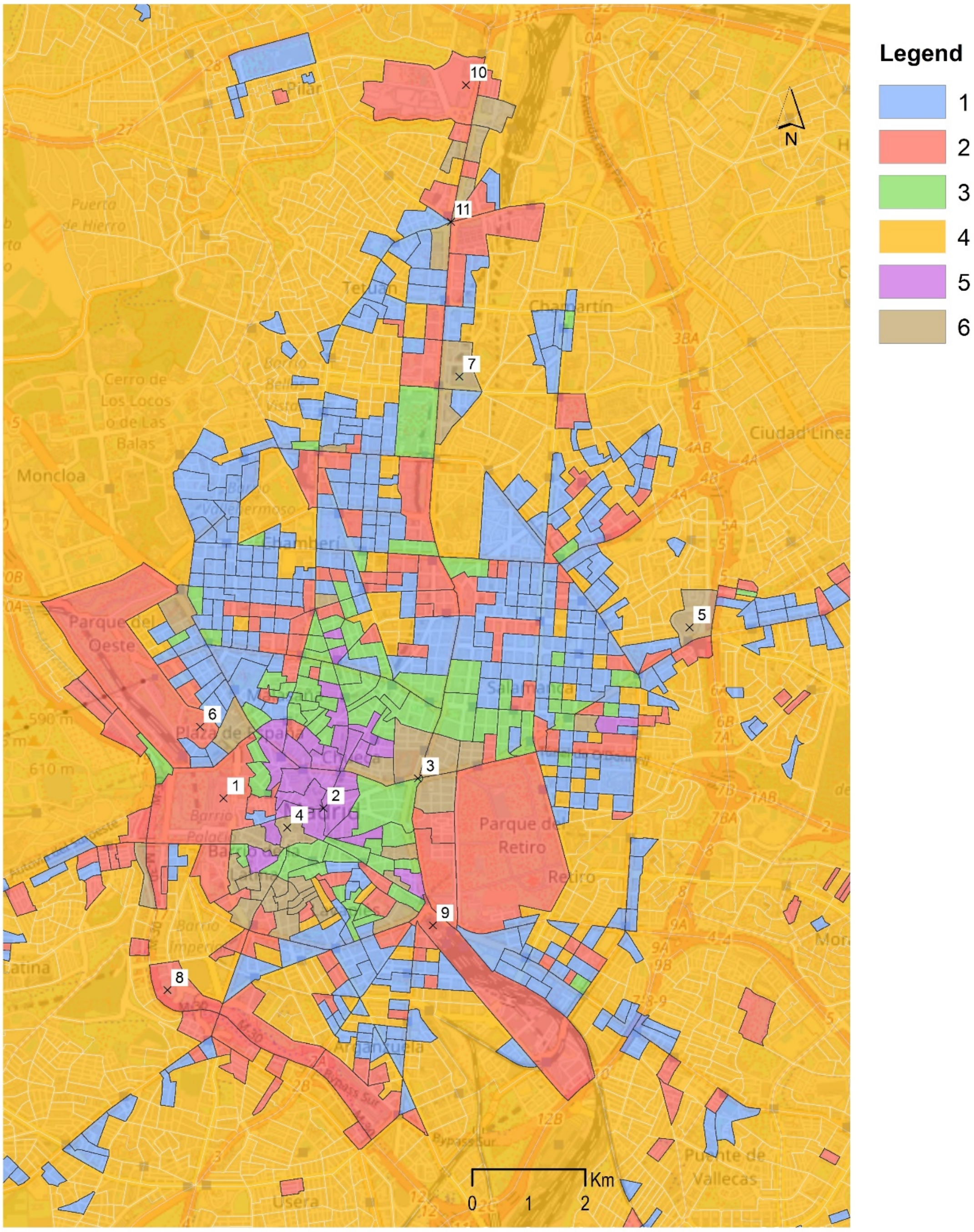

Figure 6: Groups of tracts according to tourist activities

## Analysis of spatial autocorrelation

Spatial autocorrelation allows the data from each census tract to be analysed not in isolation (as in previous sub-section) but in relation to the data in the census tracts in its environment. Using the IDW (inverse distance weighted) procedure with a distance threshold of 500 m, Global Moran's Index shows a positive spatial autocorrelation in all cases (positive Moran I values), lower for Panoramio than for Twitter and Foursquare (Table 4). Anselin Local Moran's I statistic reveals the spatial cluster distribution (significant at the 0.01 level) (Figure 7). HH census tracts (high values in a variable surrounded by high values in that same variable) tend to form a single cluster in the case of Foursquare (historic centre and Salamanca district) (Figure 7b), but several clusters in geolocated photographs (historic centre, Real Madrid stadium, Torres Kio-Cuatro Torres) (Figure 7a).

Table 4: Global Moran's I statistics (distance threshold = 500m)

| | Panoramio | Foursquare | Twitter |
|---|---|---|---|
| Global Moran's Index | 0.571 | 0.692 | 0.741 |
| z-score | 37.88 | 46.18 | 49.02 |
| p-value | 0.0000 | 0.0000 | 0.0000 |

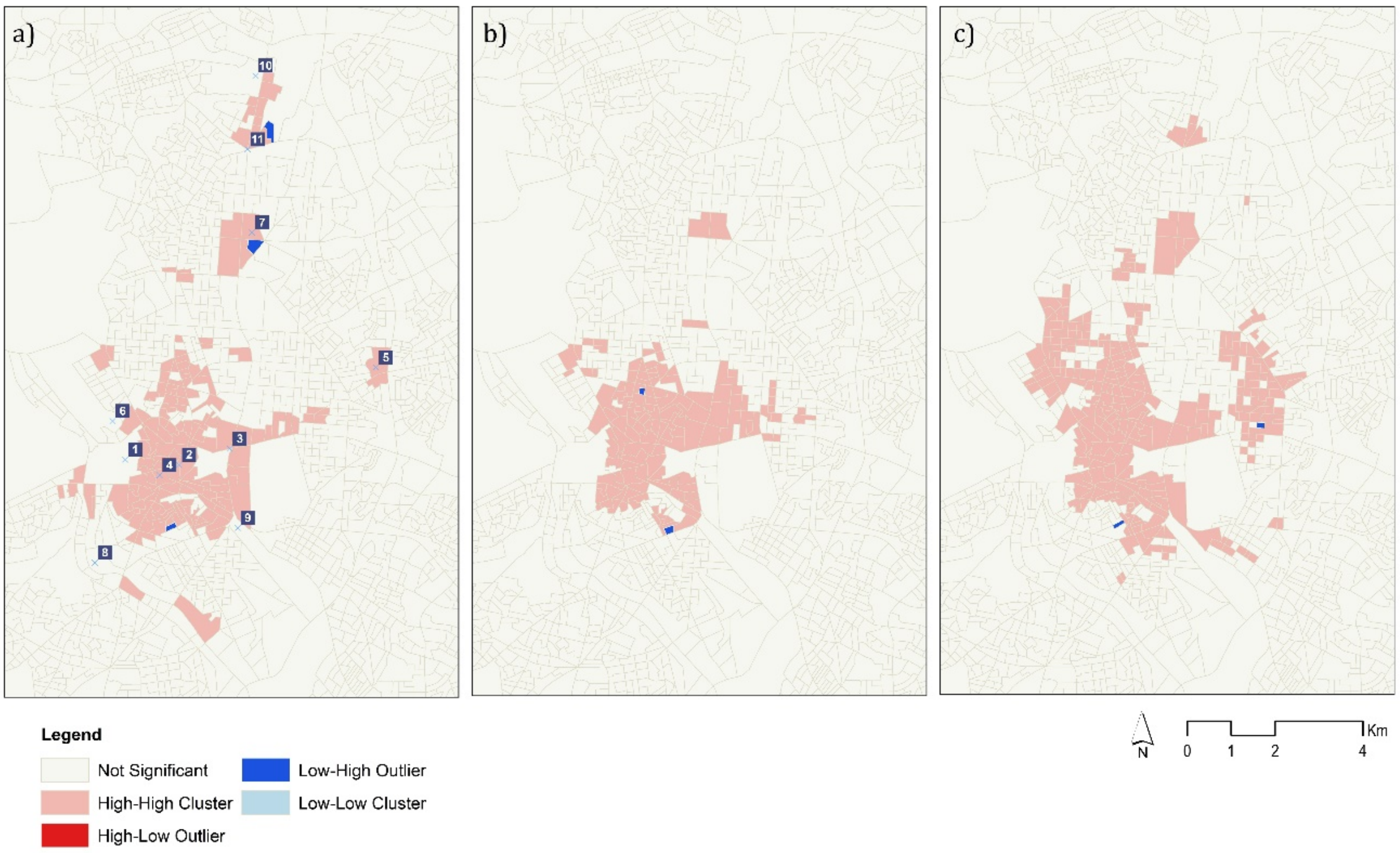

Figure 7: Results of the LISA analysis (distance threshold = 500m). a) Panoramio, b) Foursquare, c) Twitter. The references to the numbers in Figure 7a can be seen in Figure 1.

The results of the previous univariate analysis have been cross-referenced to incorporate the information from the three data sources. Figure 8 shows a classification of the census tracts considering the HH clusters of the three data sources jointly. When a census tract forms part of a HH cluster in the three sources, it indicates that in a radius of 500 m there is a high density of opportunities for sightseeing, consumption and connecting to the internet. This cross-referencing therefore gives a more complete multivariant LISA vision than would be obtained by using bivariant LISA analysis. The results show census tracts that form part of the HH cluster in the three data sources (historic centre), or in only one of them –for example, in areas specialised in sightseeing (such as Las Ventas bullring), in consumption (Salamanca district) or in both (AZCA shopping and business centre). In general, the resulting map shows how tourist specialisation tends to increase radiating outwards from the historic centre towards the periphery: the census tracts in the centre have a mixed character (multifunctional), and are surrounded by others which generally have two activities, while the most peripheral tend to specialise in one activity.

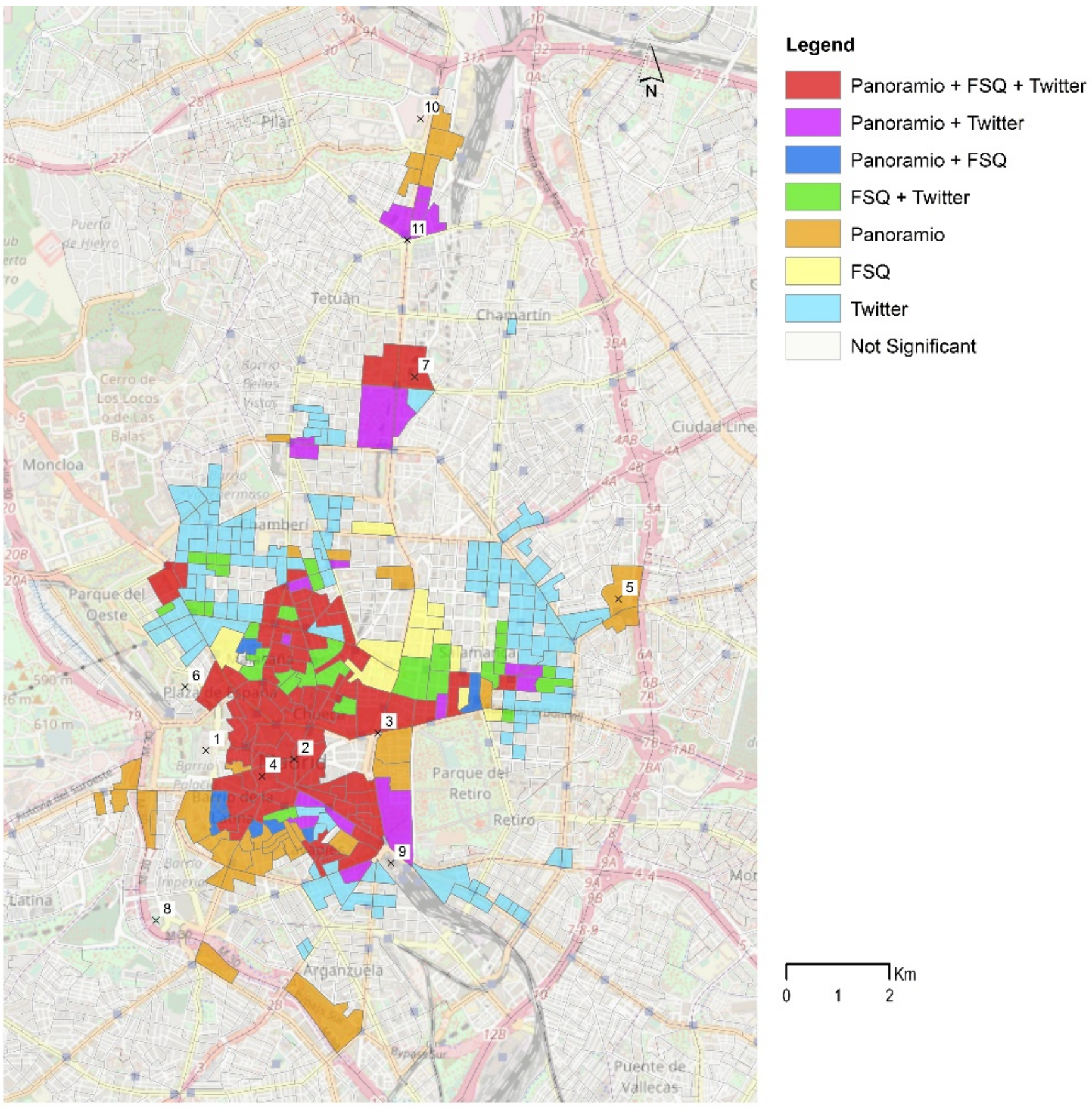

Figure 8: Typology combining the HH cluster of the three data sources

## 6. CONCLUSIONS

The work tracks the digital footprint of urban tourists using several data sources and taking the city of Madrid as the study area. Three data sources are used to identify the places where tourists carry out their activities: geolocated photographs from the Panoramio platform (sightseeing), Foursquare check-ins (consumption), and interaction on the social network Twitter (being connected). Unlike other papers, this article not only counts the density of digital footprints (for example, number of photographs or tweets) but the density of tourists. This mitigates problems of possible bias (compulsive photographers, Foursquare users or tweeters) and allows the three data sources to be compared.

The findings reveal that it is not sufficient to use a single data source to understand the spatial distribution of tourists in cities, as tourists engage in different activities in different spaces. Not surprisingly, the three data sources show a high tourist density in the historic centre, where there is a significant concentration of monuments, shops, hotels, restaurants and others. Tourists' digital footprint also extends throughout other areas of the city through photographs (football stadiums, bullring, large parks), with Foursquare check-ins (particularly in areas specialised in restaurants and shopping), and with tweets. The use of Twitter seems to be related to the location of accommodations, since Twitter data are spatially disperse covering areas with high offer of accommodation and temporally concentrated between 18 and 21 hours.

The bivariant correlation analysis reveals little similarity between the distributions of tourist density on Panoramio and Twitter –on the one hand–, and Panoramio and Foursquare –on the other–, but a certain similarity between the spatial patterns of tourists according to Foursquare and Twitter. In a second step, cluster analysis was applied to classify the city's tourist areas based on the intensity of the digital footprint left by tourists as being more oriented to sightseeing, consumption or Internet connection. Finally, the spatial autocorrelation analysis shows tourist activity by considering each census sector, not in an isolated way, but in relation to the census tracts in its surroundings. The global analysis confirms that Panoramio reveals the lowest spatial autocorrelation (with several spatial clusters). Areas allowing a wide range of different activities can be differentiated from those with a more specialised nature by combining the results of the three data sources. The degree of specialisation tends to increase from the historic centre radiating outwards towards the periphery.

As in other papers that use Big Data, there is also an underlying problem of bias. Most tourists do not upload their photos to photo-sharing communities like Panoramio, and some do not even take photographs. Photographs do not always properly reflect all the monuments in the city, due to the prohibition against taking photographs in some monuments, and particularly in museums. Many tourists do not use social networks like Twitter or Foursquare and only a small proportion of Twitter users send geolocated tweets. The source bias is unquestionably difficult to identify and correct. In this paper the bias has been mitigated by working with the density of tourists rather than with the density of their footprints (photographs, Foursquare check-ins or tweets), and avoids counting the same tourist several times –this is especially important in the case of compulsive users. In addition, by comparing different data sources we consider different tourist activities, and this partly offsets the bias caused by working with only one source.

Knowledge of the spatial distribution of urban tourists is extremely important for public policies. Thus for example in spaces with a high prevalence of tourists, the local authorities may envisage actions to improve the experience of the tourists, such as creating pedestrian-only streets or widening pavements, extending public spaces with free WiFi, locating new tourist information points, among others. The spatial distribution of the photographs taken by tourists shows there are spaces with a high tourist potential but which are still under exploited (for example, the Madrid Río axis). The results may also be important for pinpointing new opportunities for business for the private sector, for example by identifying areas with economies for locating retail tourism where there are still opportunities for expansion.

**Acknowledgments**.-